\documentclass[11pt,reqno]{amsart}
\allowdisplaybreaks[4]
\usepackage{graphicx} 
\usepackage{epsfig}
\usepackage{amssymb}
\usepackage{amsmath}
\usepackage{cite}

\newcommand{\be}{\begin{equation}}
\newcommand{\ee}{\end{equation}}
\newcommand{\bea}{\begin{eqnarray}}
\newcommand{\eea}{\end{eqnarray}}
\newcommand{\ba}{\begin{array}}
\newcommand{\ea}{\end{array}}
\newcommand{\bean}{\begin{eqnarray*}}
\newcommand{\eean}{\end{eqnarray*}}

\newcommand{\pa}{\partial}

\begin{document}

\title{Recursion operators for KP, mKP and Harry-Dym Hierarchies}
\author{Jipeng Cheng $^1$, Lihong Wang$^2$, Jingsong He $^{2*}$ }
\dedicatory { $^1$Department of Mathematics, USTC, Hefei, 230026
Anhui, P.\ R.\
China \\
$^2$ Department of Mathematics, NBU, Ningbo, 315211 Zhejiang, P.\
R.\ China }

\thanks{$^*$ Corresponding author: hejingsong@nbu.edu.cn, jshe@ustc.edu.cn}
\begin{abstract}
In this paper, we give a unified construction of the recursion
operators from the Lax representation for three integrable
hierarchies: Kadomtsev-Petviashvili (KP), modified
Kadomtsev-Petviashvili (mKP) and Harry-Dym under $n$-reduction. This
shows a new inherent relationship between them. To illustrate our
construction, the recursion operator are calculated explicitly for
$2$-reduction and $3$-reduction.
\\
{\bf AMS classification(2010):} 35Q53, 37K10, 37K40. 
\end{abstract}
\maketitle
\section{Introduction}
The recursion operator $\Phi$, firstly presented by P.J.
Olver\cite{olver1977}, plays a key role (see
\cite{olver1993,blaszak1998,dorfman1993} and references therein) in
the study of the integrable system. For single integrable evolution
equation, it always owns infinitely many commuting symmetries and
bi-Hamiltonian structures\cite{olver1993,blaszak1998,dorfman1993}
which the recursion operator can link. As for an integrable
hierarchy, the higher flows can be generated from the lower flow
with the help of the recursion operator, which offers a natural way
to construct the whole integrable hierarchy from a single seed
system (see \cite{olver1993,blaszak1998,dorfman1993} and references
therein). By now, much work has been done on the recursion operator.
For example, the construction of the recursion for a given
integrable system\cite{boiti,sf1988,fs1988,
strampp,oevel1990,xu1992,sokolov1999,blaszak,gurses2001,zk,euler2004,ab2006,sokolov2008},
and the properties of the recursion
operator\cite{zk1984,swna,swpd,gurses2002,sergyeyev,bk}. In general,
the recursion operator has non-local term. So it is a highly
non-trivial problem to understand the locality of higher order
symmetries and higher order flows generated by recursion
operator\cite{sergyeyev,olversokolov}. In this paper, we shall focus
on the construction of the recursion operator and explain the
locality of their higher flows although the recursion operator
associated is non-local.

The main object that we will investigate is three interesting
integrable hierarchies, i.e.  Kadomtsev-Petviashvili (KP), modified
Kadomtsev-Petviashvili (mKP) and Harry-Dym
hierarchies\cite{kiso,oevel1993}, which are corresponding to the
decompositions of the algebra $g$ of pseudo-differential operators
\begin{equation}\label{algebradecom}
    g:=\{\sum_{i\ll\infty}u_i\pa^i\}=\{\sum_{i\geq
    k}u_i\pa^i\}\oplus\{\sum_{i<
    k}u_i\pa^i\}:=g_{\geq k}\oplus g_{< k}
\end{equation}
for $k=0,1,2$ respectively, where $u_i$ are the functions of
$t=(t_1=x,t_2,\cdots)$ and $\pa=\pa_{x}$. The algebraic
multiplication of $\pa^i$ with the multiplication operator $u$ are
defined by
\begin{equation}\label{multiplicaiton}
    \pa^iu=\sum_{j\geq 0}C_i^ju^{(j)}\pa^{i-j},\quad i\in \mathbb{Z},
\end{equation}
where $u^{(j)}=\frac{\pa^j u}{\pa x^j}$, with
$$C_i^j=\frac{i(i-1)\cdots(i-j+1)}{j!}.$$
In fact, $g_{\geq k}$ and $g_{< k}$ are the sub-Lie algebra of $g$:
$[g_{\geq k},g_{\geq k}]\subset g_{\geq k}$ and $[g_{< k},g_{<
k}]\subset g_{< k}$
 when $k=0,1,2$. The projections of $L=\sum_i u_i\pa^i\in g$ to $g_{\geq k}$ and $ g_{<
 k}$ are
\begin{equation}\label{projection}
    L_{\geq k}=\sum_{i\geq k} u_i\pa^i,\quad L_{< k}=\sum_{i<
    k}u_i\pa^i.
\end{equation}
Then according to the famous Adler-Kostant-Symes scheme\cite{adler},
the following commuting Lax equations\cite{kiso,oevel1993} on $g$
can be constructed
\begin{equation}\label{laxequation}
    L_{t_m}=[(L^m)_{\geq k},L],
\end{equation}
where $k=0,1,2$ are corresponding to KP, mKP and Harry-Dym
hierarchies respectively, with the Lax operator $L$ given by
\begin{equation}\label{laxthreeforms}
    L = \left\{ {\begin{array}{*{20}{c}}
   {\partial  + {u_2}{\partial ^{ - 1}} + {u_3}{\partial ^{ - 2}} +  \cdots \quad k = 0},  \\
   {\partial  + {u_1} + {u_2}{\partial ^{ - 1}} +  \cdots \quad \quad k = 1},  \\
   {{u_0}\partial  + {u_1} + {u_2}{\partial ^{ - 1}} +  \cdots \quad k = 2}. \\
\end{array}} \right.
\end{equation}
For simplicity, we rewrite (\ref{laxthreeforms}) in a unified
form\cite{kiso,oevel1993}
\begin{eqnarray}
L&=&\sum_{l\geq0}u_l\pa^{1-l},\label{laxoperator}
\end{eqnarray}
i.e.
\begin{equation}\label{laxoperatornewform}
    L=\sum_{l=0}^{1-k}u_l\pa^{1-l}+\sum_{l\geq2-k}u_l\pa^{1-l},
\end{equation}
and let
\begin{eqnarray}
 B_m=(L^m)_{\geq k}, \quad L^m=\sum_{j\leq m} a_j(m)\pa^j=\sum_{j\leq m}
 \pa^jb_j(m).\label{abm}
\end{eqnarray}
Then (\ref{laxequation}) becomes into
\begin{equation}\label{introductionlaxequation}
    L_{t_m}=[B_m,L].
\end{equation}
These three integrable hierarchies have been studied intensively in
literatures\cite{ohta,kw1993,oevel1993,kupershmidt}, which contain
the following well-known $2+1$ dimensional equations
\begin{eqnarray*}
&&k=0:\quad 4u_{2tx}=(u_{2xxx}+12u_2u_{2x})_x+3u_{2yy},\quad \quad\quad\quad\quad \quad\quad\quad\quad \quad\quad\quad(\rm{KP})\\
&&k=1:\quad4u_{1tx}=(u_{1xxx}-6u_1^2u_{1x})_x+3u_{1yy}+6u_{1x}u_{1y}+6u_{1xx}\pa^{-1}u_{1y},\quad (\rm{mKP})\\
&&k=2:\quad4u_{0t}=u_0^3u_{0xxx}-3\frac{1}{u_0}(u_0^2\pa^{-1}(\frac{1}{u_0})_y)_y,\quad\quad\quad\quad\quad
\quad\quad\quad\quad \quad\quad(\rm{Harry-Dym})
\end{eqnarray*}
where we have set $t_2=y,t_3=t$.

There are some inherent relationships discovered among these three
integrable hierarchies. For example, their flow equations are
defined by a unified Lax equation (\ref{introductionlaxequation})
although their $B_m$ are different, their Hamiltonian structure is
given by a general way, i.e. $r$-matrix method\cite{kw1993}, and
there exists an interesting link among them in the flow equations
and gauge transformations\cite{oevel1993}. So it is very natural to
ask whether there exists a unified way to deal with their recursion
operators, which is just our central aim of this paper. For the KP
hierarchy, W.Strampp \& W.Oevel\cite{oevel1990} and V.V.Sokolov
\emph{et al} \cite{sokolov1999} separately developed a general
method to construct the recursion operator by the Lax representation
(\ref{introductionlaxequation}). V.V.Sokolov
\emph{et al}\cite{sokolov1999,blaszak,gurses2001,zk} used an
important ansatz $\widetilde{B}=\mathcal{P}B_n+R$ that relates $B_n$
operator for different $n$, where $\mathcal{P}$ is some operator
that commutes with the $L$ operator and $R$ is the remainder. While,
W.Strampp \& W.Oevel derived a general expression (see eq.(47) of
reference\cite{oevel1990}) for the recursion operators of the KP
hierarchy under n-reduction starting from Lax equations. In this paper, we will use W.Strampp \& W.Oevel's method. However,
their method is not applicable to get a similar and compact formula
for the mKP and Harry-Dym hierarchies due to following two
observations: 1) $(L^m)_{<0}=\sum_{j<0}\partial^j b_j(m)$ for the KP
hierarchy, but $(L^m)_{<1}=\sum_{j<1}\partial^j b_j(m)+ \sum_{j=1}^m
b_j^{(j)}(m)$ for the mKP hierarchy,
$(L^m)_{<2}=\sum_{j<2}\partial^j b_j(m)+ \sum_{j=2}^m
\left(b_j^{(j)}(m)+ j b_j^{(j-1)}(m)\pa \right) $ for the Harry-Dym
hierarchy. 2)  It is not affirmative to get a compact form of the
flow equations of  mKP hierarchy and Harry-Dym hierarchy as eq.(6)
and eq.(17) of reference \cite{oevel1990} for the KP hierarchy
because of the second summation terms in the last two cases of 1).

 In this paper, to further find inherent relations between above three hierarchies, we shall improve
W.Strampp \& W.Oevel  method (use $a_j(m)$ only ) and give a unified
construction of the recursion operators from the Lax representation for three integrable hierarchies: KP, mKP and
Harry-Dym under $n$-reduction (see eq.(\ref{nreductioncondition})).
There are two advantages in our construction:
1) it is easy to explain why nonlocal recursion operators produce local flows, since the L.H.S. of
(\ref{introductionlaxequation}) only produces the differential polynomials of $u_i$, thus the flow
equations of (\ref{introductionlaxequation}) are naturally local;
2) a formula of the recursion operator for arbitrary $n$-reduction are derived, which shows the existence
of recursion operators for the three kinds of integrable hierarchies, and provides a constructive way to get
recursion operators for higher order reductions although the calculation is not an easy task.

This paper is organized as follows. In Section 2, we rewrite the
unified Lax equations (\ref{introductionlaxequation}) into matrix
forms in terms of $a_j(m)$ under $n$-reduction. Then, we devote
Section 3 to deriving the formulas of the recursion
operators for the three integrable hierarchies. At last, we consider
the applications of the formulas of the recursion operators
and check the correctness of the formulas.

\section{Lax Equations}
In this section, we will rewrite the Lax equations
(\ref{introductionlaxequation}) into matrix forms in terms of
$a_j(m)$ under $n$-reduction. For this, we start from the $m$th
power of $L$, that is
\begin{equation}\label{npowerlax}
    L^m=\sum_{j\leq m} a_j(m)\pa^j.
\end{equation}
Thus
\begin{eqnarray}
(L^m)_{\geq k}&=&\sum_{j=k}^m a_j(m)\pa^j,\label{positivepartnpowerlax}\\
(L^m)_{< k}&=&\sum_{j< k} a_j(m)\pa^j.\label{negativepartnpowerlax}
\end{eqnarray}
Note that, the Lax dynamics equation (\ref{laxequation}) can be
rewritten into
\begin{equation}\label{anotherformlaxequation}
    L_{t_m}=[L,(L^m)_{<k}].
\end{equation}

We first derive the flow equations for the coordinates $u_i$. After
inserting (\ref{negativepartnpowerlax}) into
(\ref{anotherformlaxequation}), we find
\begin{eqnarray*}
L(L^m)_{< k}-(L^m)_{<k}L&=&\sum_{l\geq
0}\sum_{j<k}(u_l\pa^{1-l}a_j(m)\pa^j-a_j(m)\pa^ju_l\pa^{1-l})\\
&=&\sum_{l\geq 0}\sum_{j<k}\sum_{p\geq
0}(C_{1-l}^pu_la_j^{(p)}(m)-C_j^pa_j(m)u_l^{(p)})\pa^{1-l+j-p}\\
&=&\sum_{q\geq 0}\sum_{l=
0}^q\sum_{j<k}(C_{1-l}^{q-l}u_la_j^{(q-l)}(m)-C_j^{q-l}a_j(m)u_l^{(q-l)})\pa^{1-q+j}\\
&=&\sum_{r\geq 1-k}\sum_{j=-r}^{k-1}\sum_{l=
0}^{j+r}(C_{1-l}^{j+r-l}u_la_j^{(j+r-l)}(m)-C_j^{j+r-l}a_j(m)u_l^{(j+r-l)})\pa^{1-r}\\
&=&\sum_{r\geq 2-k}\sum_{j=1-k}^{r}\sum_{l=
0}^{r-j}(C_{1-l}^{r-j-l}u_la_{-j}^{(-j+r-l)}(m)-C_{-j}^{-j+r-l}a_{-j}(m)u_l^{(-j+r-l)})\pa^{1-r}
\end{eqnarray*}
According to (\ref{laxoperatornewform}), we know
\begin{equation}\label{laxtn}
    L_{t_m}=\sum_{l\geq2-k}u_{l,t_m}\pa^{1-l}.
\end{equation}
So by comparing (\ref{laxtn}) with $L(L^m)_{< k}-(L^m)_{<k}L$, we
obtain
\begin{equation}\label{explicitlaxequaion}
    u_{r,t_m}=\sum_{j=1-k}^{r}O_{r,j}a_{-j}(m),\quad
    r=2-k,3-k,\cdots,
\end{equation}
with $O_{r,j}$ given by
\begin{equation}\label{aexpression}
O_{r,j}=\sum_{l=
0}^{r-j}(C_{1-l}^{r-j-l}u_l\pa^{r-j-l}-C_{-j}^{r-j-l}u_l^{(r-j-l)}).
\end{equation}
In particular, we find
$$O_{r,r}=0,\quad O_{r,r-1}=u_0\pa-(1-r)u_{0x}.$$
Notice that from (\ref{npowerlax}), we find $a_j(m)$ can be uniquely
determined by $u_i$, that is,
\begin{equation}\label{aurelation}
    a_s(m)=mu_0^{m-1}u_{m-s}+f_{sm}(u_0,...,u_{m-s-1}),
\end{equation}
where $f_{sm}$ are the differential polynomials in
$u_0,...,u_{m-s-1}$. After substituting (\ref{aurelation}) into
(\ref{explicitlaxequaion}), we obtains a series of evolution
equations for $u_i$. These flow equations are all local because
$a_s(m)$ are the differential polynomials of $u_i$.

We next consider the so-called $n$-reduction, that is, we impose the
constraints below on the Lax operator $L$ ,
\begin{equation}\label{nreductioncondition}
    L^n=(L^n)_{\geq k}.
\end{equation}
Under the constraints above, $a_s(n)=0$ for $s<k$. Hence from
(\ref{aurelation}), we can express $u_j$ for $j>n-k$ in terms of
$(u_{2-k},u_{3-k},\cdots,u_{n-k})$. Thus only $n-1$ coordinates
$(u_{2-k},u_{3-k},\cdots,u_{n-k})$ are independent, which are in
one-to-one correspond with
$(a_k(n),a_{k+1}(n),\cdots,a_{k+n-2}(n))$. For example, under the
$2$-reduction, only $u_{2-k}$ is independent, then the flow equation
(\ref{explicitlaxequaion}) implies the following $1+1$ dimensional
equations, \\
for $k=0$
\begin{eqnarray}
u_{2t_3}&=&\frac{1}{4}u_{2xxx}+3u_2u_{2x},\label{kdv}\\
u_{2t_5} &=& \frac{15}{2}u_2^2u_{2
x}+\frac{5}{4}u_2u_{2xxx}+\frac{5}{2}u_{2xx}u_{2x}+\frac{1}{16}u_{2xxxxx},\label{5kdv}
\end{eqnarray}
for $k=1$
\begin{eqnarray}
u_{1t_3}&=& \frac{1}{4}u_{1xxx}-\frac{3}{2}u_1^2u_{1x},\label{mkdv}\\
u_{1t_5} &=&
\frac{15}{8}u_1^4u_{1x}-\frac{5}{8}u_{1xxx}u_1^2+\frac{1}{16}u_{1xxxxx}-\frac{5}{8}u_{1x}^3-\frac{5}{2}u_1u_{1xx}u_{1x},\label{5mkdv}
\end{eqnarray}
for $k=2$
\begin{eqnarray}
u_{0t_3}&=&\frac{1}{4}u_0^3u_{0xxx},\label{dym}\\
u_{0t_5}&=&\frac{1}{32}u_0^3(10u_0u_{0xx}u_{0xxx}+5u_{0xxx}u_{0x}^2+10u_0u_{0xxxx}u_{0x}+2u_0u_{0xxxxx}).\label{5dym}
\end{eqnarray}

After the preparation above, under $n$-reduction we can at last
rewrite the Lax equations (\ref{introductionlaxequation}) into
matrix forms in terms of $a_j(m)$. For this, we denote
\begin{eqnarray}
U(n)&=& (u_{2-k},u_{3-k},\cdots,u_{n-k})^t,\nonumber\\
A(n,m)&=&(a_{-1+k}(m),a_{-2+k}(m),\cdots,a_{-n+1+k}(m))^t,\nonumber\\
O(n) &= &\left( {\begin{array}{*{20}{c}}
   {{O_{2 - k,1 - k}}} & 0 &  \cdots  & 0  \\
   {{O_{3 - k,1 - k}}} & {{O_{3 - k,2 - k}}} &  \cdots  & 0  \\
    \vdots  &  \vdots  &  \ddots  &  \vdots   \\
   {{O_{n - k,1 - k}}} & {{O_{n - k,2 - k}}} &  \cdots  & {{O_{n - k,n - 1 - k}}}  \\
\end{array}} \right),\label{symbols}
\end{eqnarray}
where $t$ denotes the transpose of the matrix, then we can rewrite
(\ref{explicitlaxequaion}) into
\begin{equation}\label{matrixlaxequation}
    U(n)_{t_m}=O(n)A(n,m).
\end{equation}
It is trivial to know that all of the flow equations in $U(n)_{t_m}$
are local, including those in $ U(n)_{t_{m+jn}}$.

\section{Recursion Formulas}
In this section, we will construct the recursion operator. To do
this, we have to first obtain a recursion formula relating $A(n,m)$
and $A(n,m+n)$ under $n$-reduction constraint, that is, we try to
seek an operator $R(n)$, s.t. $A(n,m+n)=R(n)A(n,m)$.

For this, we consider the relation $L^{m+n}=L^mL^n=L^nL^m$. Assuming
$n$-reduction, we find
\begin{eqnarray}
(L^mL^n)_{< k}&=&(\sum_{j\leq k-1}\sum_{l=k}^n
a_j(m)\pa^ja_l(n)\pa^l)_{< k}\nonumber\\
&=&(\sum_{j\leq k-1}\sum_{l=k}^n\sum_{p\geq 0}
C_j^pa_j(m)a_l^{(p)}(n)\pa^{l+j-p})_{<  k}\nonumber\\
&=&(\sum_{j\leq k-1}\sum_{q\leq n}\sum_{l=max(k,q)}^n
C_j^{l-q}a_j(m)a_l^{(l-q)}(n)\pa^{q+j})_{<  k}\nonumber\\
&=&\sum_{j\leq k-1}\sum_{q=j+1-k}^n\sum_{l=max(k,q)}^n
C_{j-q}^{l-q}a_{j-q}(m)a_l^{(l-q)}(n)\pa^{j}.\label{mnlaxoperator}
\end{eqnarray}
Comparing with
\begin{equation}\label{mplusnlaxoperator}
    (L^{m+n})_{<k}=\sum_{j\leq k-1} a_j(m+n)\pa^j,
\end{equation}
we find
\begin{equation}\label{mnrelationlax}
    a_j(m+n)=\sum_{q=j+1-k}^nP_{jq}(n)a_{j-q}(m),\quad j\leq k-1,
\end{equation}
with
\begin{equation}\label{pmatrixexpression}
    P_{jq}(n)=\sum_{l=max(k,q)}^n
C_{j-q}^{l-q}a_l^{(l-q)}(n),\quad j\leq k-1,\quad
q=j+1-k,j+2-k,\cdots,n.
\end{equation}
In particular, we have
\begin{equation}\label{somepmatrix}
P_{jn}(n)=u_0^n,\quad P_{j,n-1}=a_{n-1}(n)+(j-n+1)a_{n}(n)_{x}.
\end{equation}

We next introduce the $(n-1)\times(n-1)$-matrix $S(n)$ and the
$(n-1)\times n$-matrix $T(n)$
\begin{eqnarray}
 S(n)& =& \left( {\begin{array}{*{20}{c}}
   {{P_{k - 1,0}}(n)} & {{P_{k - 1,1}}(n)} &  \cdots  & {{P_{k - 1,n - 2}}(n)}  \\
   {{P_{k - 2, - 1}}(n)} & {{P_{k - 2,0}}(n)} &  \cdots  & {{P_{k - 2,n - 3}}(n)}  \\
    \vdots  &  \vdots  &  \ddots  &  \vdots   \\
   {{P_{ - n + 1 + k, - n + 2}}(n)} & {{P_{ - n + 1 + k, - n + 3}}(n)} &  \cdots  & {{P_{ - n + 1 + k,0}}(n)}  \\
\end{array}} \right), \label{smatrix}\\
 T(n) &= &\left( {\begin{array}{*{20}{c}}
   {{P_{k - 1,n - 1}}(n)} & {{P_{k - 1,n}}(n)} & 0 &  \cdots  & 0  \\
   {{P_{k - 2,n - 2}}(n)} & {{P_{k - 2,n - 1}}(n)} & {{P_{k - 2,n}}(n)} &  \cdots  & 0  \\
    \vdots  &  \vdots  &  \vdots  &  \ddots  &  \vdots   \\
   {{P_{ - n + 1 + k,1}}(n)} & {{P_{ - n + 1 + k,2}}(n)} & {{P_{ - n + 1 + k,3}}(n)} &  \cdots  & {{P_{ - n + 1 + k,n}}(n)}  \\
\end{array}} \right).\label{tmatrix}
\end{eqnarray}
So (\ref{mnrelationlax}) for $j=-1+k,-2+k,\cdots,-n+1+k$ can be
written into
\begin{equation}\label{chubujielun}
    A(n,m+n)=S(n)A(n,m)+T(n)(a_{-n+k}(m),a_{-n-1+k}(m),\cdots,a_{-2n+1+k}(m))^t.
\end{equation}

So now the only thing that we need to do is to express
$(a_{-n+k}(m),a_{-n-1+k}(m),\cdots,a_{-2n+1+k}(m))^t$ in terms of
$A(n,m)$. For this, we will use the relation $L^mL^n=L^nL^m$.
\begin{eqnarray*}
(L^nL^m)_{<k}&=&(\sum_{l=k}^n \sum_{j\leq
k-1}a_l(n)\pa^la_j(m)\pa^j)_{<k}\\
&=&(\sum_{l=k}^n \sum_{j\leq
k-1}\sum_{s= 0}^lC_l^sa_l(n)a_j^{(s)}(m)\pa^{j+l-s})_{<k}\\
&=&(\sum_{j\leq
k-1}\sum_{\mu=0}^n \sum_{s= max(k-\mu,0)}^{n-\mu}C_{s+\mu}^sa_{s+\mu}(n)a_j^{(s)}(m)\pa^{j+\mu})_{<k}\\
&=&\sum_{j\leq k-1}\sum_{\mu=0}^n \sum_{s=
max(k-\mu,0)}^{n-\mu}C_{s+\mu}^sa_{s+\mu}(n)a_{j-\mu}^{(s)}(m)\pa^{j},
\end{eqnarray*}
Comparing with (\ref{mplusnlaxoperator}), we obtain
\begin{equation}\label{anothermnrelation}
    a_j(m+n)=\sum_{\mu=0}^n Q_\mu(n)a_{j-\mu}(m),\quad j\leq k-1,
\end{equation}
with
\begin{equation}\label{qexpression}
    Q_\mu(n)=\sum_{s=
max(k-\mu,0)}^{n-\mu}C_{s+\mu}^sa_{s+\mu}(n)\pa^s,\quad 0\leq
\mu\leq n.
\end{equation}
In particular,
\begin{equation}\label{someqexpression}
 Q_n(n)=u_0^n,\quad Q_{n-1}(n)=na_n(n)\pa+a_{n-1}(n).
\end{equation}
Thus using $(\ref{mnrelationlax})=(\ref{anothermnrelation})$, we
obtain
\begin{eqnarray*}
 &&\left( {\begin{array}{*{20}{c}}
   {{a_{ - 1 + k}}(m + n)}  \\
   {{a_{ - 2 + k}}(m + n)}  \\
    \vdots   \\
   {{a_{ - n + 1 + k}}(m + n)}  \\
   {{a_{ - n + k}}(m + n)}  \\
\end{array}} \right) \\
 && = \left( {\begin{array}{*{20}{c}}
   {{P_{ - 1 + k,0}}(n)} & {{P_{ - 1 + k,1}}(n)} &  \cdots  & {{P_{ - 1 + k,n - 3}}(n)} & {{P_{ - 1 + k,n - 2}}(n)}  \\
   {{P_{ - 2 + k, - 1}}(n)} & {{P_{ - 2 + k,0}}(n)} &  \cdots  & {{P_{ - 2 + k,n - 4}}(n)} & {{P_{ - 2 + k,n - 3}}(n)}  \\
    \vdots  &  \vdots  &  \ddots  &  \vdots  &  \vdots   \\
   {{P_{ - n + 1 + k, - n + 2}}(n)} & {{P_{ - n + 1 + k, - n + 3}}(n)} &  \cdots  & {{P_{ - n + 1 + k, - 1}}(n)} & {{P_{ - n + 1 + k,0}}(n)}  \\
   {{P_{ - n + k, - n + 1}}(n)} & {{P_{ - n + k, - n + 2}}(n)} &  \cdots  & {{P_{ - n + k, - 2}}(n)} & {{P_{ - n + k, - 1}}(n)}  \\
\end{array}} \right)\left( {\begin{array}{*{20}{c}}
   {{a_{ - 1 + k}}(m)}  \\
   {{a_{ - 2 + k}}(m)}  \\
    \vdots   \\
   {{a_{ - n + 2 + k}}(m)}  \\
   {{a_{ - n + 1 + k}}(m)}  \\
\end{array}} \right) \\
  &&+ \left( {\begin{array}{*{20}{c}}
   {{P_{ - 1 + k,n - 1}}(n)} & 0 &  \cdots  & 0 & 0  \\
   {{P_{ - 2 + k,n - 2}}(n)} & {{P_{ - 2 + k,n - 1}}(n)} &  \cdots  & 0 & 0  \\
    \vdots  &  \vdots  &  \ddots  &  \vdots  &  \vdots   \\
   {{P_{ - n + 1 + k,1}}(n)} & {{P_{ - n + 1 + k,2}}(n)} &  \cdots  & {{P_{ - n + 1 + k,n - 1}}(n)} & 0  \\
   {{P_{ - n + k,0}}(n)} & {{P_{ - n + k,1}}(n)} &  \cdots  & {{P_{ - n + k,n - 2}}(n)} & {{P_{ - n + k,n - 1}}(n)}  \\
\end{array}} \right)\left( {\begin{array}{*{20}{c}}
   {{a_{ - n + k}}(m)}  \\
   {{a_{ - n - 1 + k}}(m)}  \\
    \vdots   \\
   {{a_{ - 2n + 2 + k}}(m)}  \\
   {{a_{ - 2n + 1 + k}}(m)}  \\
\end{array}} \right) \\
 && + u_0^n\left( {\begin{array}{*{20}{c}}
   {{a_{ - n - 1 + k}}(m)}  \\
   {{a_{ - n - 2 + k}}(m)}  \\
    \vdots   \\
   {{a_{ - 2n + 1 + k}}(m)}  \\
   {{a_{ - 2n + k}}(m)}  \\
\end{array}} \right) = \left( {\begin{array}{*{20}{c}}
   {{Q_0}(n)} & {{Q_1}(n)} &  \cdots  & {{Q_{n - 3}}(n)} & {{Q_{n - 2}}(n)}  \\
   0 & {{Q_0}(n)} &  \cdots  & {{Q_{n - 4}}(n)} & {{Q_{n - 3}}(n)}  \\
    \vdots  &  \vdots  &  \ddots  &  \vdots  &  \vdots   \\
   0 & 0 &  \cdots  & 0 & {{Q_0}(n)}  \\
   0 & 0 &  \cdots  & 0 & 0  \\
\end{array}} \right)\left( {\begin{array}{*{20}{c}}
   {{a_{ - 1 + k}}(m)}  \\
   {{a_{ - 2 + k}}(m)}  \\
    \vdots   \\
   {{a_{ - n + 2 + k}}(m)}  \\
   {{a_{ - n + 1 + k}}(m)}  \\
\end{array}} \right) \\
  &&+ \left( {\begin{array}{*{20}{c}}
   {n{a_n}(n)\partial  + {a_{n - 1}}(n)} & 0 &  \cdots  & 0 & 0  \\
   {{Q_{n - 2}}(n)} & {n{a_n}(n)\partial  + {a_{n - 1}}(n)} &  \cdots  & 0 & 0  \\
    \vdots  &  \vdots  &  \ddots  &  \vdots  &  \vdots   \\
   {{Q_1}(n)} & {{Q_2}(n)} &  \cdots  & {n{a_n}(n)\partial  + {a_{n - 1}}(n)} & 0  \\
   {{Q_0}(n)} & {{Q_1}(n)} &  \cdots  & {{Q_{n - 2}}(n)} & {n{a_n}(n)\partial  + {a_{n - 1}}(n)}  \\
\end{array}} \right) \\
  &&\times \left( {\begin{array}{*{20}{c}}
   {{a_{ - n + k}}(m)}  \\
   {{a_{ - n - 1 + k}}(m)}  \\
    \vdots   \\
   {{a_{ - 2n + 2 + k}}(m)}  \\
   {{a_{ - 2n + 1 + k}}(m)}  \\
\end{array}} \right) + u_0^n\left( {\begin{array}{*{20}{c}}
   {{a_{ - n - 1 + k}}(m)}  \\
   {{a_{ - n - 2 + k}}(m)}  \\
    \vdots   \\
   {{a_{ - 2n + 1 + k}}(m)}  \\
   {{a_{ - 2n + k}}(m)}  \\
\end{array}} \right).
\end{eqnarray*}
So if we denote $M(n)$ and $N(n)$ as $n\times n$ and $n\times (n-1)$
respectively, that is
\begin{eqnarray*}
 &&M(n) =  \\
 &&\left( {\begin{array}{*{20}{c}}
   { - n{a_n}(n)\partial  - (n - k){a_n}{{(n)}_x}} & 0 &  \cdots  & 0  \\
   {{D_{ - 2 + k,n - 2}}(n)} & { - n{a_n}(n)\partial  - (1 - k + n){a_n}{{(n)}_x}} &  \cdots  & 0  \\
    \vdots  &  \vdots  &  \ddots  &  \vdots   \\
   {{D_{ - n + k,0}}(n)} & {{D_{ - n + k,1}}(n)} &  \cdots  & { - n{a_n}(n)\partial  - (2n - 1 - k){a_n}{{(n)}_x}}  \\
\end{array}} \right),\\
&& N(n) = \left( {\begin{array}{*{20}{c}}
   {{D_{ - 1 + k,0}}(n)} & {{D_{ - 1 + k,1}}(n)} &  \cdots  & {{D_{ - 1 + k,n - 3}}(n)} & {{D_{ - 1 + k,n - 2}}(n)}  \\
   {{P_{ - 2 + k, - 1}}(n)} & {{D_{ - 2 + k,0}}(n)} &  \cdots  & {{D_{ - 2 + k,n - 4}}(n)} & {{D_{ - 2 + k,n - 3}}(n)}  \\
    \vdots  &  \vdots  &  \ddots  &  \vdots  &  \vdots   \\
   {{P_{ - n + 1 + k, - n + 2}}(n)} & {{P_{ - n + 1 + k, - n + 3}}(n)} &  \cdots  & {{P_{ - n + 1 + k, - 1}}(n)} & {{D_{ - n + 1 + k,0}}(n)}  \\
   {{P_{ - n + k, - n + 1}}(n)} & {{P_{ - n + k, - n + 2}}(n)} &  \cdots  & {{P_{ - n + k, - 2}}(n)} & {{P_{ - n + k, - 1}}(n)}  \\
\end{array}} \right),
\end{eqnarray*}
with $D_{js}=P_{j,s}(n)-Q_{s}(n)$, then we have
\begin{eqnarray}
&&M(n)(a_{-n+k}(m),a_{-n-1+k}(m),\cdots,a_{-2n+2+k}(m),a_{-2n+1+k}(m))^t\nonumber\\
    &&=-N(n)(a_{-1+k}(m),a_{-2+k}(m),\cdots,a_{-n+2+k}(m),a_{-n+1+k}(m))^t.\label{jinyibujielun}
\end{eqnarray}
Since $M(n)$ is invertible, we can solve
$(a_{-n+k}(m),a_{-n-1+k}(m),\cdots,a_{-2n+2+k}(m),a_{-2n+1+k}(m))^t$
from (\ref{jinyibujielun}) and then insert into (\ref{chubujielun}).

So we get
\begin{equation}\label{recursionformula}
    A(n,m+n)=R(n)A(n,m)
\end{equation}
with
\begin{equation}\label{recursionoperator}
    R(n)=S(n)-T(n)M(n)^{-1}N(n).
\end{equation}
If we set
\begin{equation}\label{newrecursionoperator}
    \Phi(n)=O(n)R(n)O(n)^{-1},
\end{equation}
then we can easily find
\begin{eqnarray}
 U(n)_{t_{m+jn}}&=&O(n)A(n,m+jn)\nonumber\\
 &=&O(n)R(n)A(n,m+(j-1)n)\nonumber\\
 &=&O(n)R(n)O(n)^{-1}O(n)A(n,m+(j-1)n)\nonumber\\
&=&\Phi(n)U(n)_{t_{m+(j-1)n}}\nonumber\\
&&\cdots\nonumber\\
&=&\Phi(n)^jU(n)_{t_{m}}.\label{recursionscheme}
\end{eqnarray}

\textbf{Remark:} Note that the recursion operator
(\ref{newrecursionoperator}) is nonlocal, but it does not generate
the nonlocal higher flow equations, because in our cases, all the
flow equations in (\ref{matrixlaxequation}) are local and the
recursion operator (\ref{newrecursionoperator}) is just extracted
from these local flow equations.

\section{Applications}
In this section, we give some examples for the applications of the
formula (\ref{newrecursionoperator}) for the recursion operator.
Here we only consider $2$-reduction and $3$-reduction.

2-REDUCTION

For $k=0$ case, one calculates
\begin{eqnarray*}
&& {a_0}(2) = 2{u_2},\quad O(2) = \partial ,\quad S(2) = {a_0}(2),\quad T(2) = (0,1), \\
&& M(2) = \left( {\begin{array}{*{20}{c}}
   { - 2\partial } & 0  \\
   { - {\partial ^2}} & { - 2\partial }  \\
\end{array}} \right),\quad M{(2)^{ - 1}} = \left( {\begin{array}{*{20}{c}}
   { - \frac{1}{2}{\partial ^{ - 1}}} & 0  \\
   {\frac{1}{4}} & { - \frac{1}{2}{\partial ^{ - 1}}}  \\
\end{array}} \right),\quad  \\
&& N(2) = \left( {\begin{array}{*{20}{c}}
   { - {\partial ^2}}  \\
   { - {a_0}{{(2)}_x}}  \\
\end{array}} \right),\quad R(2) = \frac{1}{4}{\partial ^2} + 2{u_2} - {\partial ^{ -
1}}{u_{2x}}.
\end{eqnarray*}
So the recursion operator\cite{oevel1990,sokolov1999} is
\begin{equation}\label{kdvresursion}
    \Phi(2)=\frac{1}{4}{\partial ^2} + 2{u_2} + {u_{2x}}{\partial ^{ -
1}}.
\end{equation}

Beginning from $u_{2t_1}=u_{2x}$, we find
\begin{eqnarray*}
u_{2t_3}&=& \Phi(2)u_{2t_1}=\frac{1}{4}u_{2xxx}+3u_2u_{2x},\\
u_{2t_5} &=& \Phi(2)u_{2t_3}= \frac{15}{2}u_2^2u_{2
x}+\frac{5}{4}u_2u_{2xxx}+\frac{5}{2}u_{2xx}u_{2x}+\frac{1}{16}u_{2xxxxx}.
\end{eqnarray*}
Note that in the reference\cite{oevel1990}, W.Strampp \& W.Oevel
also calculate the $M(2)$ and $N(2)$, but they are different from
here. This is because we only use $a_j(m)$.

For $k=1$ case, one has
\begin{eqnarray*}
&& {a_1}(2) = 2{u_1},\quad O(2) = \partial ,\quad S(2) = 0,\quad T(2) = ({a_1}(2),1), \\
&& M(2) = \left( {\begin{array}{*{20}{c}}
   { - 2\partial } & 0  \\
   { - {a_1}{{(2)}_x} - {a_1}(2) - {\partial ^2}} & { - 2\partial }  \\
\end{array}} \right),\quad M{(2)^{ - 1}} = \left( {\begin{array}{*{20}{c}}
   { - \frac{1}{2}{\partial ^{ - 1}}} & 0  \\
   {\frac{1}{4}{a_1}(2){\partial ^{ - 1}} + \frac{1}{4}} & { - \frac{1}{2}{\partial ^{ - 1}}}  \\
\end{array}} \right),\quad  \\
&& N(2) = \left( {\begin{array}{*{20}{c}}
   { - {a_1}(2) - {\partial ^2}}  \\
   0  \\
\end{array}} \right),\quad R(2) = \frac{1}{4}{\partial ^2} - \frac{1}{4}{a_1}(2){\partial ^{ -
1}}{a_1}(2)\partial.
\end{eqnarray*}
The corresponding recursion operator \cite{olver1977} is
\begin{equation}\label{mkdvrecursion}
\Phi(2)=\frac{1}{4}{\partial ^2} - u_1^2- {u_{1x}}{\partial ^{ -
1}}u_1.
\end{equation}
Thus, from $u_{1t_1}=u_{1x}$, we know
\begin{eqnarray*}
u_{1t_3}&=& \Phi(2)u_{1t_1}=\frac{1}{4}u_{1xxx}-\frac{3}{2}u_1^2u_{1x},\\
u_{1t_5} &=&
\Phi(2)u_{1t_3}=\frac{15}{8}u_1^4u_{1x}-\frac{5}{8}u_{1xxx}u_1^2+\frac{1}{16}u_{1xxxxx}-\frac{5}{8}u_{1x}^3-\frac{5}{2}u_1u_{1xx}u_{1x}.
\end{eqnarray*}

At last, for $k=2$, we have
\begin{eqnarray*}
&& O(2) = {u_0}\partial  - {u_{0x}} = u_0^2\partial u_0^{ - 1},\quad O{(2)^{ - 1}} = {u_0}{\partial ^{ - 1}}u_0^{ - 2},\quad S(2) = 0,\quad T(2) = (0,u_0^2), \\
&& M(2) = \left( {\begin{array}{*{20}{c}}
   { - 2{u_0}^2\partial } & 0  \\
   { - u_0^2{\partial ^2}} & { - 2{u_0}({u_0}\partial  + {u_{0x}})}  \\
\end{array}} \right),\quad M{(2)^{ - 1}} = \left( {\begin{array}{*{20}{c}}
   { - \frac{1}{2}{\partial ^{ - 1}}u_0^{ - 2}} & 0  \\
   {\frac{1}{4}u_0^{ - 1}{\partial ^{ - 1}}{u_0}\partial u_0^{ - 2}} & { - \frac{1}{2}u_0^{ - 1}{\partial ^{ - 1}}u_0^{ - 1}}  \\
\end{array}} \right),\quad  \\
&& N(2) = \left( {\begin{array}{*{20}{c}}
   { - {u_0}^2{\partial ^2}}  \\
   0  \\
\end{array}} \right),\quad R(2) = \frac{1}{4}{u_0}{\partial ^{ - 1}}{u_0}{\partial
^3}.
\end{eqnarray*}
Therefore the recursion operator \cite{dym} is
\begin{equation}\label{harrydymrecursion}
    \Phi(2)=\frac{1}{4}u_0^3\pa^3u_0\pa^{-1}u_0^{-2}.
\end{equation}
So
\begin{eqnarray*}
u_{0t_1}&=&0,\\
u_{0t_3}&=&\Phi(2)u_{0t_1}=\frac{1}{4}u_0^3u_{0xxx},\\
u_{0t_5}&=&\Phi(2)u_{0t_3}=\frac{1}{32}u_0^3(10u_0u_{0xx}u_{0xxx}+5u_{0xxx}u_{0x}^2+10u_0u_{0xxxx}u_{0x}+2u_0u_{0xxxxx}).
\end{eqnarray*}

Obviously, all of above soliton equations are consistent with flow
equations of (\ref{kdv})-(\ref{5dym}), which shows the validity of
the explicit recursion operators
(\ref{kdvresursion})-(\ref{harrydymrecursion}).

3-REDUCTION

$k=0$ case
\begin{eqnarray*}
&&O(3)=\left(\begin{array}{cc}
       \pa & 0 \\
       0 & \pa
     \end{array}\right),
\quad S(3)=\left(\begin{array}{cc}
a_0(3)-a_{1}(3)_x & a_1(3) \\
-a_{0}(3)_x+a_{1}(3)_{xx} & a_0(3)-2a_{1}(3)_x
\end{array}\right),\\
&&
 T(3) = \left( {\begin{array}{*{20}{c}}
   0 & 1 & 0  \\
   {{a_1}(3)} & 0 & 1  \\
\end{array}} \right),\quad M(3) = \left( {\begin{array}{*{20}{c}}
   { - 3\partial } & 0 & 0  \\
   { - 3{\partial ^2}} & { - 3\partial } & 0  \\
   { - 3{a_1}{{(3)}_x} - {a_1}(3)\partial  - {\partial ^3}} & { - 3{\partial ^2}} & { - 3\partial }  \\
\end{array}} \right), \\
 &&M{(3)^{ - 1}} = \left( {\begin{array}{*{20}{c}}
   { - \frac{1}{3}{\partial ^{ - 1}}} & 0 & 0  \\
   {\frac{1}{3}} & { - \frac{1}{3}{\partial ^{ - 1}}} & 0  \\
   { - \frac{2}{9}\partial  + \frac{1}{3}{a_1}(3){\partial ^{ - 1}} - \frac{2}{9}{\partial ^{ - 1}}{a_1}(3)} & {\frac{1}{3}} & { - \frac{1}{3}{\partial ^{ - 1}}}  \\
\end{array}} \right), \\
 &&N(3) = \left( {\begin{array}{*{20}{c}}
   { - {a_1}{{(3)}_x} - {a_1}(3)\partial  - {\partial ^3}} & { - 3{\partial ^2}}  \\
   { - {a_0}{{(3)}_x} + {a_1}{{(3)}_{xx}}} & { - 2{a_1}{{(3)}_x} - {a_1}(3)\partial  - {\partial ^3}}  \\
   {{a_0}{{(3)}_{xx}} - {a_1}{{(3)}_{xxx}}} & { - 2{a_0}{{(3)}_x} + 3{a_1}{{(3)}_{xx}}}  \\
\end{array}} \right)
\end{eqnarray*}
with $a_1(3)=3u_2, a_0(3)=3u_3+3u_{2x}$. Then by (\ref{recursionoperator}) and (\ref{newrecursionoperator}), the recursion operator is given by
\begin{equation}\label{30recursion}
    \Phi(3)=\left(\begin{array}{cc}
                    \Phi_{11} & \Phi_{12} \\
                    \Phi_{21} & \Phi_{22}
                  \end{array}    \right)
\end{equation}
where
\begin{eqnarray*}
\Phi_{11}&=&\frac{2}{3}\pa a_0(3)\pa^{-1}-\frac{1}{3}\pa a_1(3)_x\pa^{-1}+\frac{1}{3}\pa a_1(3)+\frac{1}{3}\pa^3+\frac{1}{3}a_0(3)-\frac{1}{3}a_1(3)_x,\\
\Phi_{12}&=&\frac{1}{3}\pa a_1(3)\pa^{-1}+\frac{2}{3}\pa^2+\frac{1}{3}a_1(3),\\
\Phi_{21}&=&-\frac{1}{3}\pa a_0(3)_x\pa^{-1}+\frac{1}{3}\pa a_1(3)_{xx}\pa^{-1}-\frac{2}{9}\pa^3a_1(3)\pa^{-1}-\frac{2}{9}\pa^4-\frac{2}{9}a_1(3)\pa a_1(3)\pa^{-1}\\
&&-\frac{2}{9}a_1(3)\pa^2-\frac{1}{3}a_0(3)_x+\frac{1}{3}a_1(3)_{xx},\\
\Phi_{22}&=&\frac{1}{3}\pa a_0(3)\pa^{-1}-\frac{1}{3}\pa a_1(3)_{x}\pa^{-1}-\frac{1}{3}\pa^3-\frac{2}{3}a_1(3)\pa+\frac{1}{3}\pa a_1(3)+\frac{2}{3}a_0(3)-a_1(3)_x.
\end{eqnarray*}
We have checked the action of recursion operator (\ref{30recursion}) on the $t_1$ flow to $t_4$ flow, that is,
\begin{eqnarray*}
 {\left( {\begin{array}{*{20}{c}}
   {{u_2}}  \\
   {{u_3}}  \\
\end{array}} \right)_{{t_1}}} &= &{\left( {\begin{array}{*{20}{c}}
   {{u_2}}  \\
   {{u_3}}  \\
\end{array}} \right)_x} \\
 {\left( {\begin{array}{*{20}{c}}
   {{u_2}}  \\
   {{u_3}}  \\
\end{array}} \right)_{{t_4}}} &= &\Phi (3){\left( {\begin{array}{*{20}{c}}
   {{u_2}}  \\
   {{u_3}}  \\
\end{array}} \right)_{{t_1}}} \\
  &= &\left( {\begin{array}{*{20}{c}}
   {4{u_3}{u_{2x}} + 4{u_2}{u_{3x}} + 2{u_2}{u_{2xx}} + 2u_{_{2x}}^2 + \frac{2}{3}{u_{3xxx}} + \frac{1}{3}{u_{2xxxx}}}  \\
   { - 2{u_{3x}}{u_{2x}} - 2{u_2}{u_{3xx}} - 2{u_2}{u_{2xxx}} - \frac{1}{3}{u_{3xxxx}} - \frac{2}{9}{u_{2xxxxx}} - 4u_2^2{u_{2x}} + 4{u_3}{u_{3x}} - 4{u_{2xx}}{u_{2x}}}  \\
\end{array}} \right)
\end{eqnarray*}
$k=1$ case,
\begin{eqnarray*}
 &&O(3)= \left( {\begin{array}{*{20}{c}}
   \partial  & 0  \\
   0 & \partial   \\
\end{array}} \right),\quad S(3) = \left( {\begin{array}{*{20}{c}}
   0 & {{a_1}(3) - {a_2}{{(3)}_x}}  \\
   0 & { - {a_1}{{(3)}_x} + {a_2}{{(3)}_{xx}}}  \\
\end{array}} \right), \\
 &&T(3) = \left( {\begin{array}{*{20}{c}}
   {{a_2}(3)} & 1 & 0  \\
   {{a_1}(3) - 2{a_2}{{(3)}_x}} & {{a_2}(3)} & 1  \\
\end{array}} \right), \\
 &&M(3) = \left( {\begin{array}{*{20}{c}}
   { - 3\partial } & 0 & 0  \\
   { - 2{a_2}{{(3)}_x} - 2{a_2}(3)\partial  - 3{\partial ^2}} & { - 3\partial } & 0  \\
   { - 2{a_1}{{(3)}_x} + 3{a_2}{{(3)}_{xx}} - {a_1}(3)\partial  - {a_2}(3){\partial ^2} - {\partial ^3}} & { - 3{a_2}{{(3)}_x} - 2{a_2}(3)\partial  - 3{\partial ^2}} & { - 3\partial }  \\
\end{array}} \right), \\
 &&M{(3)^{ - 1}} = \left( {\begin{array}{*{20}{c}}
   { - \frac{1}{3}{\partial ^{ - 1}}} & 0 & 0  \\
   A & { - \frac{1}{3}{\partial ^{ - 1}}} & 0  \\
   B & C & { - \frac{1}{3}{\partial ^{ - 1}}}  \\
\end{array}} \right), \\
 &&N(3) = \left( {\begin{array}{*{20}{c}}
   { - {a_1}(3)\partial  - {a_2}(3){\partial ^2} - {\partial ^3}} & { - {a_2}{{(3)}_x} - 2{a_2}(3)\partial  - 3{\partial ^2}}  \\
   0 & { - {a_1}{{(3)}_x} + {a_2}{{(3)}_{xx}} - {a_1}(3)\partial  - {a_2}(3){\partial ^2} - {\partial ^3}}  \\
   0 & {{a_1}{{(3)}_{xx}} - {a_2}{{(3)}_{xxx}}}  \\
\end{array}} \right),
\end{eqnarray*}
where
\begin{eqnarray*}
a_1(3)&=&3u_2+3u_1^2+3u_{1x},\quad a_2(3)=3u_1,\\
 A&=&\frac{2}{9}a_2(3)\pa ^{-1}+\frac{1}{3},\quad C=\frac{1}{3}a_2(3)\pa ^{-1}-\frac{1}{9}\pa ^{-1}a_2(3)+\frac{1}{3},\\
 B&=&\frac{2}{9}a_1(3)\pa ^{-1}-\frac{1}{9}\pa ^{-1}a_1(3)-\frac{5}{9}\pa a_2(3)\pa ^{-1}-\frac{2}{9}a_2^2(3)\pa ^{-1}+\frac{2}{27}\pa^{-1}a_2(3)\pa a_2(3) \pa ^{-1}\\
 &&\frac{1}{3}a_2(3)-\frac{1}{9}\pa^{-1}a_2(3)\pa-\frac{2}{9}\pa.
\end{eqnarray*}
Then according to (\ref{recursionoperator}) and (\ref{newrecursionoperator}), we get the recursion operator
\begin{equation}\label{31recursion}
    \Phi(3)=\left(\begin{array}{cc}
                    \Phi_{11} & \Phi_{12} \\
                    \Phi_{21} & \Phi_{22}
                  \end{array}    \right)
\end{equation}
where
\begin{eqnarray*}
\Phi_{11}&=&-\frac{1}{9}\pa a_2(3)\pa^{-1}a_1(3)-\frac{1}{9}\pa a_2(3)\pa^{-1}a_2(3)\pa+\frac{2}{9}\pa a_2(3)\pa+\frac{1}{3}\pa a_1(3)+\frac{1}{3}\pa^3,\\
\Phi_{12}&=&\frac{2}{3}\pa a_1(3)\pa^{-1}-\frac{1}{3}\pa a_2(3)_x\pa^{-1}-\frac{1}{9}\pa a_2(3)^2\pa^{-1}-\frac{1}{9}\pa a_2(3)\pa^{-1}a_2(3)+\frac{2}{3}\pa^2,\\
\Phi_{21}&=&-\frac{1}{9}\pa a_1(3)\pa-\frac{1}{9}a_1(3)\pa^2-\frac{1}{9}\pa a_1(3)\pa^{-1}a_1(3)-\frac{1}{9}a_1^2(3)-\frac{1}{9}\pa a_1(3)\pa^{-1}a_2(3)\pa\\
&&-\frac{1}{9}a_1(3)a_2(3)\pa+\frac{1}{9}\pa^2 a_2(3)\pa^{-1}a_1(3)-\frac{1}{9}\pa^2 a_2(3)\pa+\frac{1}{9}\pa^2 a_2(3)\pa^{-1}a_2(3)\pa\\
&&+\frac{2}{27}a_2(3)\pa a_2(3)\pa^{-1}a_1(3)+\frac{2}{27}a_2(3)\pa a_2(3)\pa^{-1}a_2(3)\pa+\frac{2}{27}a_2(3)\pa a_2(3)\pa\\
&&-\frac{1}{9}a_2(3)\pa a_1(3)-\frac{1}{9}a_2(3)\pa a_2(3)\pa-\frac{1}{9}a_2(3)\pa^3 -\frac{2}{9}\pa^2a_1(3)-\frac{2}{9}\pa^4,\\
\Phi_{22}&=&-\frac{1}{3}\pa a_1(3)_x\pa^{-1}+\frac{1}{3}\pa a_2(3)_{xx}\pa^{-1}-\frac{1}{9}\pa a_1(3)a_2(3)\pa^{-1}-\frac{1}{9}\pa a_1(3)\pa^{-1}a_2(3)\\
&&-\frac{1}{3}\pa a_1(3)-\frac{1}{9}a_1(3)a_2(3)_x\pa^{-1}-\frac{1}{3}a_1(3)a_2(3)-\frac{1}{3}a_1(3) \pa+\frac{1}{9}\pa^2 a_2^2(3)\pa^{-1}\\
&&+\frac{1}{9}\pa^2 a_2(3)\pa^{-1}a_2(3)-\frac{1}{9}\pa^2 a_2(3)+\frac{2}{27}a_2(3)\pa a_2^2(3)\pa^{-1}+\frac{2}{27}a_2(3)\pa a_2(3)\pa^{-1}a_2(3)\\
&&-\frac{1}{9}a_2(3)a_1(3)_x\pa^{-1}+\frac{1}{9}a_2(3)a_2(3)_{xx}\pa^{-1}-\frac{1}{9}a_2^2(3)\pa-\frac{1}{9}a_2(3)\pa^{2}+\frac{1}{3}\pa a_2(3)_x\\
&&+\frac{1}{3} a_1(3)\pa-\frac{1}{9}a_2(3)\pa a_2(3)_{x}\pa^{-1}-\frac{2}{9}\pa^2a_2(3)_x\pa^{-1}-\frac{1}{3}\pa^3.
\end{eqnarray*}
With the recursion operator (\ref{31recursion}), we can get the $t_4$ flow from $t_1$ flow,
\begin{eqnarray*}
u_{1t_4}&=& \frac{1}{3}u_{1xxxx}+\frac{2}{3}u_{2xxx}+2u_{2x}u_{1x}-\frac{4}{3}u_{1x}u_1^3+2u_2u_{1xx}+4u_2u_{2x}+\frac{2}{3}u_1u_{1xxx}+\frac{8}{3}u_{1x}u_{1xx},\\
u_{2t_4} &=& -4u_1u_{1xx}u_{1x}-8u_1u_2u_{2x}-4u_1^2u_2u_{1x}-4u_1u_2u_{1xx}-4u_1u_{2x}u_{1x}-4u_2u_{1x}^2-\frac{1}{3}u_{2xxxx}\\
&&-\frac{2}{9}u_{1xxxxx}-4u_2^2u_{1x}-\frac{2}{3}u_{1xxx}u_1^2-\frac{4}{3}u_1^3u_{2x}-2u_2u_{1xxx}-\frac{2}{3}u_1u_{1xxxx}-\frac{8}{3}u_{1x}u_{1xxx}\\
&&-\frac{2}{3}u_{1}u_{2xxx}-2u_{2xx}u_{1x}-\frac{10}{3}u_{2x}u_{1xx}-2u_2u_{2xx}-2u_{1xx}^2-2u_{2x}^2-\frac{4}{3}u_{1x}^3.
\end{eqnarray*}

$k=2$ case,
\begin{eqnarray*}
 &&O(3) = \left( {\begin{array}{*{20}{c}}
   {u_0^2\partial u_0^{ - 1}} & 0  \\
   { - {u_{1x}}} & {{u_0}\partial }  \\
\end{array}} \right),\quad O{(3)^{ - 1}} = \left( {\begin{array}{*{20}{c}}
   {{u_0}{\partial ^{ - 1}}u_0^{ - 2}} & 0  \\
   { {\partial ^{ - 1}}{u_{1x}}{\partial ^{ - 1}}u_0^{ - 2}} & {{\partial ^{ - 1}}u_0^{ - 1}}  \\
\end{array}} \right),\quad  \\
 &&S(3) = \left( {\begin{array}{*{20}{c}}
   0 & 0  \\
   0 & 0  \\
\end{array}} \right),\quad T(3) = \left( {\begin{array}{*{20}{c}}
   {{P_{12}}} & {{P_{13}}} & 0  \\
   {{P_{01}}} & {{P_{02}}} & {{P_{03}}}  \\
\end{array}} \right), \\
 &&M(3) = \left( {\begin{array}{*{20}{c}}
   { - 3{a_3}(3)\partial  - {a_{3x}}} & 0 & 0  \\
   {{D_{01}}} & { - 3{a_3}(3)\partial  - 2{a_{3x}}} & 0  \\
   {{D_{ - 1,0}}} & {{D_{ - 1,1}}} & { - 3{a_3}(3)\partial  - 3{a_{3x}}}  \\
\end{array}} \right), \\
&& M{(3)^{ - 1}} = \left( {\begin{array}{*{20}{c}}
   { - \frac{1}{3}u_0^{ - 1}{\partial ^{ - 1}}u_0^{ - 2}} & 0 & 0  \\
   A & { - \frac{1}{3}u_0^{ - 2}{\partial ^{ - 1}}u_0^{ - 1}} & 0  \\
   B & C & { - \frac{1}{3}u_0^{ - 3}{\partial ^{ - 1}}}  \\
\end{array}} \right),\quad  \\
 &&N(3) = \left( {\begin{array}{*{20}{c}}
   {{D_{10}}} & {{D_{11}}}  \\
   0 & {{D_{00}}}  \\
   0 & 0  \\
\end{array}} \right)
\end{eqnarray*}
with
\begin{eqnarray*}
&&a_2(3)=3u_0^2(u_1+u_{0x}),\quad a_3(3)=u_0^3,\\
&&P_{12}=a_2(3)-a_3(3)_x,\quad P_{13}=a_3(3),\quad P_{01}=-a_2(3)_x+a_3(3)_{xx},\\
&&P_{12}=a_2(3)-2a_3(3)_x,\quad P_{13}=a_3(3),\\
&&D_{01}=-a_2(3)_x+a_3(3)_{xx}-2a_2(3)\pa-3a_3(3)\pa^2,\\
&&D_{-1,0}=a_2(3)_{xx}-a_3(3)_{xxx}-a_2(3)\pa^2-a_3(3)\pa^3,\\
&&D_{-1,1}=-2a_2(3)_x+3a_3(3)_{xx}-2a_2(3)\pa-3a_3(3)\pa^2,\\
&&D_{10}=-a_2(3)\pa^2-a_3(3)\pa^3,\quad D_{11}=-2a_2(3)\pa-3a_3(3)\pa^2,\\
&&D_{00}=-a_2(3)\pa^2-a_3(3)\pa^3,\\
&&A=-\frac{1}{9}u_0^{-2}\pa^{-1}u_0^{-1}D_{01}u_0^{-1}\pa^{-1}u_0^{-2},\quad C=-\frac{1}{9}u_0^{-3}\pa^{-1}D_{-1,1}u_0^{-2}\pa^{-1}u_0^{-1},\\
&&B=-\frac{1}{9}u_0^{-3}\pa^{-1}D_{-1,0}u_0^{-1}\pa^{-1}u_0^{-2}-\frac{1}{27}u_0^{-3}\pa^{-1}D_{-1,1}u_0^{-2}\pa^{-1}u_0^{-1}D_{01}u_0^{-1}\pa^{-1}u_0^{-2}.
\end{eqnarray*}
Then the recursion operator can be got with the help of (\ref{recursionoperator}) and (\ref{newrecursionoperator}).
\begin{equation}\label{32recursion}
    \Phi(3)=\left(\begin{array}{cc}
                    \Phi_{11} & \Phi_{12} \\
                    \Phi_{21} & \Phi_{22}
                  \end{array}    \right)
\end{equation}
where
\begin{eqnarray*}
\Phi_{11}&=&\frac{1}{3}u_0^2\pa u_0^{-2}P_{12}\pa^{-1}u_0^{-2}D_{10}u_0\pa^{-1} u_0^{-2}-u_0^2\pa u_0^{-1}P_{13}AD_{10}u_0\pa^{-1} u_0^{-2} \\
&&+\frac{1}{3}u_0^2\pa u_0^{-2}P_{12}\pa^{-1}u_0^{-2}D_{11}\pa^{-1}u_{1x}\pa^{-1} u_0^{-2}-u_0^2\pa u_0^{-1}P_{13}AD_{11}\pa^{-1}u_{1x}\pa^{-1} u_0^{-2}\\
&&+\frac{1}{3}u_0^2\pa u_0^{-3}P_{13}\pa^{-1}u_0^{-1}D_{00}\pa^{-1}u_{1x}\pa^{-1} u_0^{-2} ,\\
\Phi_{12}&=&\frac{1}{3}u_0^2\pa u_0^{-2}P_{12}\pa^{-1}u_0^{-2}D_{11}\pa^{-1} u_0^{-1}-u_0^2\pa u_0^{-1}P_{13}AD_{11}\pa^{-1} u_0^{-1}\\
&&+\frac{1}{3}u_0^2\pa u_0^{-3}P_{13}\pa^{-1}u_0^{-1}D_{00}\pa^{-1}u_0^{-1} ,\\
\Phi_{21}&=&-\frac{1}{3}u_{1x}u_0^{-1}P_{12}\pa^{-1}u_0^{-2}D_{10}u_0\pa^{-1} u_0^{-2}+u_{1x}P_{13}AD_{10}u_0\pa^{-1} u_0^{-2} \\
&&-\frac{1}{3}u_{1x}u_0^{-1}P_{12}\pa^{-1}u_0^{-2}D_{11}\pa^{-1}u_{1x}\pa^{-1} u_0^{-2}+u_{1x}P_{13}AD_{11}\pa^{-1}u_{1x}\pa^{-1} u_0^{-2}\\
&&-\frac{1}{3}u_{1x}u_0^{-2}P_{13}\pa^{-1}u_0^{-1}D_{00}\pa^{-1}u_{1x}\pa^{-1} u_0^{-2}+\frac{1}{3}u_{0}\pa P_{01}u_0^{-1}\pa^{-1}u_0^{-2}D_{10}u_0\pa^{-1} u_0^{-2} \\
&&-u_0\pa P_{02}A D_{10}u_0\pa^{-1}u_0^{-2}-u_0\pa P_{03}B D_{10}u_0\pa^{-1}u_0^{-2}\\
&&+\frac{1}{3}u_{0}\pa P_{01}u_0^{-1}\pa^{-1}u_0^{-2}D_{11}\pa^{-1}u_{1x}\pa^{-1} u_0^{-2}-u_0\pa P_{02}A D_{11}\pa^{-1}u_{1x}\pa^{-1} u_0^{-2}\\
&&-u_0\pa P_{03}B D_{11}\pa^{-1}u_{1x}\pa^{-1} u_0^{-2}+\frac{1}{3}u_{0}\pa P_{02}u_0^{-2}\pa^{-1}u_0^{-1}D_{00}\pa^{-1}u_{1x}\pa^{-1} u_0^{-2}\\
&&-u_0\pa P_{03}C D_{00}\pa^{-1}u_{1x}\pa^{-1}u_0^{-2},\\
\Phi_{22}&=&-\frac{1}{3}u_{1x}u_0^{-1}P_{12}\pa^{-1}u_0^{-2}D_{11}\pa^{-1} u_0^{-1}+u_{1x}P_{13}AD_{11}\pa^{-1} u_0^{-1} \\
&&-\frac{1}{3}u_{1x}u_0^{-2}P_{13}\pa^{-1}u_0^{-1}D_{00}\pa^{-1} u_0^{-1}+\frac{1}{3}u_{0}\pa P_{01}u_0^{-1}\pa^{-1}u_0^{-2}D_{11}\pa^{-1} u_0^{-1} \\
&&-u_0\pa P_{02}A D_{11}\pa^{-1}u_0^{-1}-u_0\pa P_{03}B D_{11}\pa^{-1}u_0^{-1}\\
&&+\frac{1}{3}u_{0}\pa P_{02}u_0^{-2}\pa^{-1}u_0^{-1}D_{00}\pa^{-1}u_0^{-1}-u_0\pa P_{03}C D_{00}\pa^{-1}u_0^{-1}.
\end{eqnarray*}
The $t_1$ and $t_4$ flows are as follows,
\begin{eqnarray*}
u_{0t_1}&=&u_{1t_1}=0,\\
u_{0t_4}&=& \frac{1}{3}u_0^2(-12u_{1x}u_1^2-12u_{1x}u_{0x}u_1+2u_0^2u_{1xxx}+2u_0u_{0xx}^2+u_0^2u_{0xxxx}+4u_{1x}u_0u_{0xx}+6u_{0x}u_0u_{1xx}\\
&&+2u_0u_{0xxx}u_1+4u_0u_{0xxx}u_{0x}-2u_{1x}u_{0x}^2-u_{0xx}u_{0x}^2-6u_{0xx}u_1^2-6u_{0xx}u_{0x}u_1),\\
u_{1t4} &=& -\frac{1}{9}u_0(24u_{1x}u_{0x}u_0u_{0xx}+6u_0u_{0xxx}u_1u_{0x}-30u_{1x}u_1u_0u_{0xx}-6u_{0xx}u_{0x}^3-72u_{0xx}u_{0x}u_1^2\\
&&+3u_0^3u_{1xxxx}+18u_{1x}u_0^2u_{0xxx}-36u_{0xx}u_1^3+33u_0u_{1xx}u_{0x}^2-12u_{1x}u_{0x}^3-84u_{1x}u_1u_{0x}^2\\
&&+16u_0^2u_{0xxxx}u_{0x}-18u_0u_{1xx}u_1^2+18u_{0xx}u_0^2u_{0xxx}+24u_0^2u_{1xx}u_{0xx}-42u_{0xx}u_1u_{0x}^2\\
&&-36u_1u_0u_{1x}^2+18u_{1x}u_0^2u_{1xx}-6u_1u_0u_{0xx}^2-12u_0u_{0xxx}u_1^2+22u_0u_{0xxx}u_{0x}^2+6u_0^2u_{0xxxx}u_1\\
&&-72u_{1x}u_1^3+12w_{0x}u_0u_{0xx}^2-144u_{1x}u_{0x}u_1^2+2u_0^3u_{0xxxxx}+6u_0^2u_{1xxx}u_1+24u_0^2u_{1xxx}u_{0x}).
\end{eqnarray*}
Because of the complication of the corresponding calculations, we have only checked $u_{0t_4}$ by the recursion operator (\ref{32recursion}) from the $t_1$ flow.

{\bf Remark:} From the examples above, one easily find that though these recursion operators contain the nonlocal terms, the flows generated by them are local. And also the forms of the recursion operators are much more complicated when $n$ and $k$ tend to larger.

{\bf Acknowledgments} {\small This work is supported by NSF of China
under grant number 10971109. He is also supported by the Program for
NCET under Grant No.NCET-08-0515. We sincerely thanks Professors Li
Yishen and Cheng Yi(USTC, China) for long-term support and
encouragement.}

\end{document}